\documentclass[cits]{PoS}
\usepackage[intlimits]{amsmath}
\usepackage{amssymb}
\usepackage{bbm}
\newcommand\T{\rule{0pt}{2.6ex}}
\newcommand\B{\rule[-1.2ex]{0pt}{0pt}}

\title{$D_s$ meson spectroscopy}

\ShortTitle{$D_s$ meson spectroscopy}

\author{\speaker{Daniel Mohler}\hspace{1mm}$^a$, R.~M. Woloshyn$^a$ \\
        \llap{$^a$} TRIUMF,4004 Wesbrook Mall, Vancouver, BC V6T 2A3, Canada\\
        E-mail: \email{mohler@triumf.ca},\email{rwww@triumf.ca}}

\abstract{Preliminary results are presented for the spectrum of $D_s$ mesons using the 2+1 flavor Clover-Wilson configurations made available by the PACS-CS collaboration. For the heavy quark, the Fermilab method is employed and we report on the tuning of the charm-quark hopping parameter. As our main focus, we present initial results for the spectrum of P-wave states, where previous results have been mostly from quenched calculations. As a cross-check, some calculations of the charmonium spectrum are also carried out.}

\FullConference{The XXVIII International Symposium on Lattice Field Theory, Lattice2010\\
		June 14-19, 2010\\
		Villasimius, Italy}

\begin{document}

\section{Introduction}

The spectrum of charmed-strange mesons contains a number of well established states, notably the ``S-wave'' states of quantum numbers $J^P$ $0^-$
and $1^-$, the $D_s$ and the $D_s^\star$ and the ``P-wave'' states with quantum numbers $0^+$ ($D_{s0}^\star(2317)$), $1^+$ ($D_{s1}(2460)$ and $D_{s1}^\star(2536)$) and $2^+$ ($D_{s2}^\star(2573)$). In addition there are a number of states which have been observed more recently. These are \cite{0954-3899-37-7A-075021} the $D_{s1}^\star (2710)$ the $D_{sJ}^\star(2860)$ the $D_{sJ}^\star(3040)$) and an unconfirmed state previously observed by SELEX \cite{Evdokimov:2004iy}, the $D_{sJ}^\star(2632)$. While the $D_{s1}^\star (2710)$ is commonly believed to have quantum numbers $J^P=1^-$, there are several possibilities for the other states which are not ruled out by experiment. The $D_{sJ}^\star(2860)$ has natural parity and is most often identified with a $3^-$ state, while some still argue the possibility of a $0^+$ identification \cite{vanBeveren:2009jq}. The $D_{sJ}^\star(3040)$ has unnatural parity and is commonly interpreted as either a $1^+$ or a $2^-$ state.

Lattice QCD (LQCD) provides the possibility to elucidate the spectrum without resorting to model assumptions. To reach this goal, several systematic sources of uncertainty have to be controlled. This has recently been achieved for the light-quark ground state meson and baryon spectrum (\cite{Durr:2008zz}). Dealing with heavy charm or bottom quarks on the lattice introduces complications of its own, and so far a similar precision for the spectrum of $D_s$ states has not been attained.

One common feature of calculations within various models and early LQCD calculations is that the ground states in the $0^+$ and $1^+$ channels are often found to be quite a bit heavier than the experimental resonances. This has sparked speculations about the nature of these states. Molecular or tetraquark interpretations have been suggested. Another contribution to these proceedings \cite{Gong:2010} deals with this possibility.

 In the next section our calculational setup and the tuning of the charm quark mass are described. Some first preliminary results from a limited number of configurations follow in section \ref{results}. As a cross check, the low-lying charmonium spectrum is also calculated. We conclude with a short summary and outlook.

\begin{table}[bht]
\begin{center}
\begin{tabular}{|c|c|c|c|c|c|c|c|}
\hline
 \T\B lattice size & $\beta$ & $c_{sw}^{(l)}$ & $c_{sw}^{(h)}$ & $\kappa_{u/d}$ &  $\kappa_s$  & \#configs used & \#configs total\\
\hline
\T\B $32^3\times 64$ & 1.90 & 1.715 & 1.52617 & 0.13700 & 0.13640 & 104 & 399\\
\hline
\T\B $32^3\times 64$ & 1.90 & 1.715 & 1.52493 & 0.13727 & 0.13640 & - & 400\\
\hline
\T\B $32^3\times 64$ & 1.90 & 1.715 & 1.52381 & 0.13754 & 0.13640 & - & 450\\
\hline
\T\B $32^3\times 64$ & 1.90 & 1.715 & 1.52327 & 0.13754 & 0.13660 & - & 400\\
\hline
\T\B $32^3\times 64$ & 1.90 & 1.715 & 1.52326 & 0.13770 & 0.13640 & 66 & 800\\
\hline
\T\B $32^3\times 64$ & 1.90 & 1.715 & 1.52264 & 0.13781 & 0.13640 & 104 & 198\\
\hline
\end{tabular}
\end{center}
\label{paratable}
\caption{Run parameters for the PACS-CS lattices \cite{Aoki:2008sm}. With $c_{sw}^{(h)}$ we denote the heavy quark clover term. \# configs indicates our current preliminary statistics.}
\end{table}

\section{\label{calc}Calculational setup}

Dynamical 2+1 flavor Clover-Wilson configurations generated by the PACS-CS collaboration \cite{Aoki:2008sm} are used in this work. They span sea-quark pion masses from 702 MeV down to 156 MeV with a lattice spacing of 0.0907(13) fm, as determined in \cite{Aoki:2008sm}. Table \ref{paratable} shows the parameters for the runs. The preliminary results presented here are based on a limited number of configurations as indicated in the table.

\subsection{Charm quark treatment}

To determine the mass parameter for the heavy charm quark we use the \emph{Fermilab method} \cite{ElKhadra:1996mp} as employed by the Fermilab-MILC collaboration \cite{Bernard:2010fr} for their efforts involving charm and bottom quarks. Within this approach, the charm quark hopping parameter $\kappa_c$ is tuned to the value where the spin averaged \emph{kinetic mass} $(M_{D_s}+3M_{D_s^*})/4$ assumes its physical value. In this simplest formulation the heavy quark hopping parameter $c_E=c_B=c_{sw}^{(h)}$ is set to its tadpole improved value $\frac{1}{u_0^3}$, where the average link $u_0$ is determined from the plaquette. The lattice dispersion relation takes the general form \cite{Bernard:2010fr}
\begin{align}
E(p)&=M_1+\frac{p^2}{2M_2}-\frac{a^3W_4}{6}\sum_ip_i^4-\frac{(p^2)^2}{8M_4^3}+ \dots .
\end{align}
As there are not enough points to constrain such a fit we use the following simplified fit forms:
\begin{itemize}
\item [1] neglect the term with coefficient $W_4$ and fit $M_1$, $M_2$ and $M_4$.
\item [2] fit $E^2(p)$ and neglect the $(p^2)^2$ term arising from the mismatch of $M_1$, $M_2$ and $M_4$
\begin{align}
E^2(p)&\approx M_1^2+\frac{M_1}{M_2}p^2-\frac{M_1a^3W_4}{3}\sum_i(p_i)^4 .
\end{align}
\end{itemize}

\begin{figure}[bht]
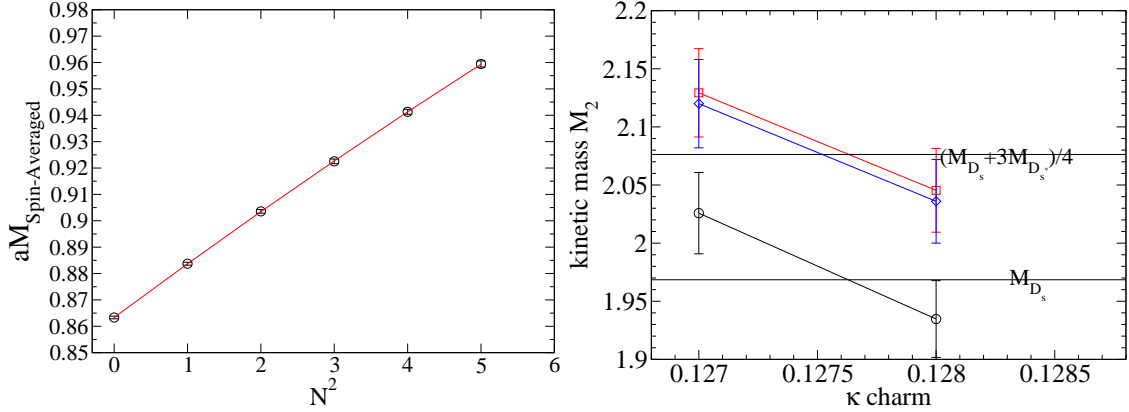

\begin{center}
\includegraphics[height=5.4cm,clip]{tuning_example_method_1.eps}
\includegraphics[height=5.4cm,clip]{kappa_tuning_results.eps}
\caption{Left-hand side: Example fit for $\kappa_c=0.128$ obtained from Method 1 (as described in the text). The resulting fit parameters are $M_1=0.86334(50)$, $M_2=0.9337(73)$ and  $M_4=0.863(28)$ leading to $\frac{M_2}{M_1}=1.0815(86)$. Right-hand side: Linear interpolation to determine the $\kappa_c$ for the spectrum calculation. Results for both the $D_s$ $0^-$ state and the spin averaged mass are shown. The red curve (squares) is a shift to account for the effect of a slightly unphysical strange quark mass.}
\label{tuning}
\end{center}
\end{figure}

The left-hand side of Fig. \ref{tuning} shows an example fit obtained from the first method. The resulting parameters are given in the caption. While the statistical errors are somewhat larger with the second method, both methods lead to reasonable fits which are compatible within statistical errors. The right-hand side of Fig.\ref{tuning} shows our final tuning results. Based on a linear interpolation, we use $\kappa_c=0.12752$ for the determination of the mass spectrum. To arrive at this value the results have been shifted to account for the somewhat unphysical strange quark mass used in the simulation.

\subsection{Simulation details \& Source construction}

Propagators are calculated for 8 source time slices on each (independent) gauge configuration. The source locations are chosen randomly within the time slice. For the calculation of the strange quark propagators the dfl\_sap\_gcr inverter from L\"uschers DDHMC package \cite{Luscher:2007se,Luscher:2007es} is employed. For the charm quark propagators the corresponding inverter without deflation is used. All error bars shown are determined with a single-elimination jackknife procedure and are of a purely statistical nature.

A matrix of interpolators is constructed in each channel and the variational method \cite{Luscher:1990ck,Michael:1985ne} used to isolate the low-lying spectrum. To this end, both Jacobi-smeared \cite{Gusken:1989ad,Best:1997qp} Gaussian sources $u_s\equiv (S \,u)_x$
\begin{align*}
S&=M \; S_0\quad\mbox{with}\quad M=\sum_{n=0}^N\kappa^nH^n ,\\
H(\vec{n},\vec{m}\,)&=\sum_{j = 1}^3
\left(U_j\left(\vec{n},0\right) \delta\left(\vec{n} + \hat{j}, \vec{m}\right)+ U_j\left(\vec{n}-\hat{j\,},0\right)^\dagger 
\delta\left(\vec{n} - \hat{j}, \vec{m}\right) 
\right),
\end{align*}
and derivative sources $W_{d_i}$
\begin{align*}
D_i(\vec{x},\vec{y})&=U_i(\vec{x},0)\delta(\vec{x}+\hat{i},\vec{y})-U_i(\vec{x}-\hat{i},0)^\dagger\delta(\vec{x}-\hat{i},\vec{y})\
,\\
W_{d_i}&=D_i\,S\ 
\end{align*}
are calculated. While we use both Gaussian and derivative sources for the charm quark, only Gaussian sources are used for the strange quark.

\section{\label{results}Results for $D_s$ mesons and charmonium}

\begin{figure}[bt]
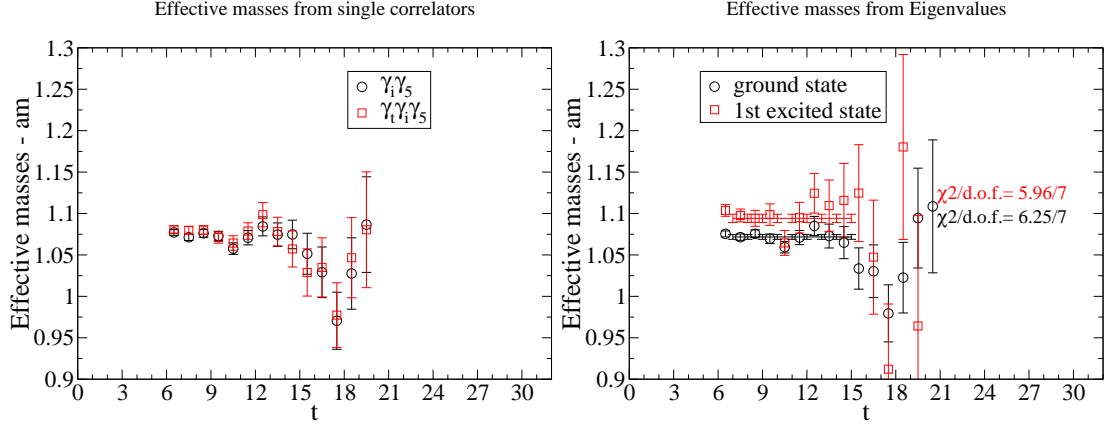

\begin{center}
\includegraphics[height=5.6cm,clip]{hl_1+_single.eps}
\includegraphics[height=5.6cm,clip]{hl_1+_mixing.eps}
\caption{Left-hand side: Diagonal correlators in the $1^+$ channel from the $\kappa$ tuning run at $\kappa_c=0.128$. The interpolators corresponding to different charge conjugation in the mass degenerate case lead to almost identical masses. Right-hand side: Using a $2\times 2$ matrix of correlartors two separate low-lying states can be identified. The mixing between the two structures enhances the observed mass splitting.}
\label{mixing}
\end{center}
\end{figure}

In the $1^+$ sector interpolators corresponding to different charge conjugation in the mass degenerate case are expected to mix. This important effect is illustrated in Fig. \ref{mixing}. While the single correlators (left-hand side of the figure) are almost degenerate, one can clearly isolate two distinct low-lying states when considering a matrix of two or more suitable interpolators (right-hand side of the figure). A similar effect is expected in the $2^-$ channel.

\begin{figure}[bt]
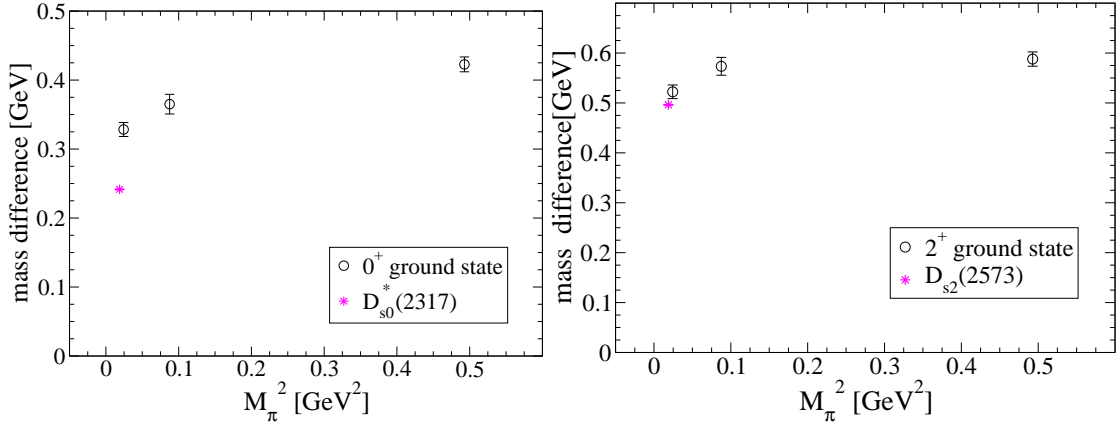

\begin{center}
\includegraphics[height=5.6cm,clip]{hl_0+_chiral_v2.eps}
\includegraphics[height=5.6cm,clip]{hl_2+_T2_chiral.eps}
\caption{Chiral behavior in the $0^+$ and $2^+$ channels. For the latter the results from the T2 lattice irreducible representation are shown. The results from the E representation are similar.}
\label{chiral_examples}
\end{center}
\end{figure}

Figure \ref{chiral_examples} shows preliminary results for the ground states in the $0^+$ and $2^+$ channels. In this figure as in all following figures mass splittings relative to the spin-averaged pole mass are plotted. The results highlight the need for simulations at light sea quark masses\footnote{Note that the detailed behavior when approaching the chiral limit is intimately related to the scale setting on the lattice.}. While our results in the $2^+$ channel show only a small deviation from the experimental value (fully consistent with the presence of non-negligible discretization effects), the results in the $0^+$ channel still suggest a ground state energy somewhat larger than the physical one. This could be attributed to volume effects, discretization effects and/or the lack of scattering states in our basis.

\begin{figure}[bt]
\begin{center}
\includegraphics[width=9cm,clip]{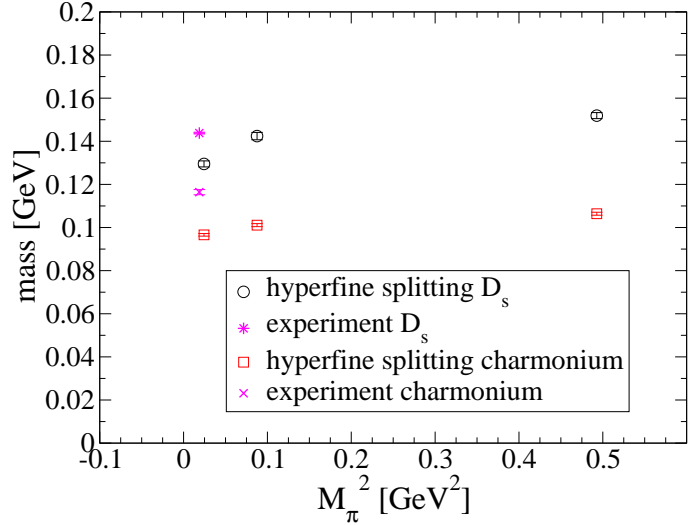}
\caption{Hyperfine splitting for $D_s$ and charmonium compared to the physical splitting. The leading discrepancy can be attributed to discretization errors.}
\label{hyperfine}
\end{center}
\end{figure}

To get an idea of the magnitude of discretization effects it is instructive to take a closer look at the ground state hyperfine splitting. Figure \ref{hyperfine} shows the results for both $D_s$ mesons and for charmonium. As expected \cite{Bernard:2010fr} within the Fermilab approach the discrepancy is smaller for $D_s$ than for charmonium, where the experimental value is underestimated by about 18 MeV. One may expect discretization effects for the P-wave states to be of similar magnitude.

\begin{figure}[bt]
\begin{center}
\includegraphics[height=5.6cm,clip]{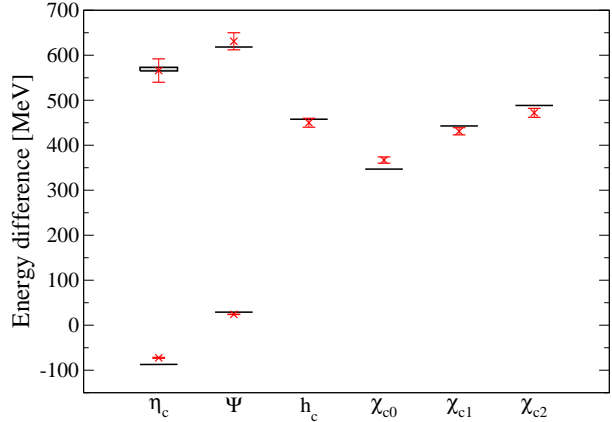}
\caption{Overview of low-lying charmonium states. The data points are from the simulation at $m_\pi=156$ MeV and the errors are statistical only. Where appropriate, the error of the experimental state is indicated by the thickness of the lines. }
\label{charm_overview}
\end{center}
\end{figure}

While we have not accumulated enough data points for a sensible extrapolation, one can get a first idea of the low-lying charmonium spectrum by comparing our simulation results at the lightest pion mass with the experimental spectrum. Such a comparison is shown in Fig. \ref{charm_overview}.

\section{Summary and outlook}

Preliminary results were presented for a simulation of the charmed-strange $D_s$ mesons and of the low-lying charmonium spectrum. As a tool of choice the variational method has been used. The calculations indicate the importance of light sea quarks for some of the states. Results for the hyperfine splitting and the low-lying charmonium spectrum show that a reasonable mass-tuning for the charm quark has been achieved. For the $D_s$ mesons with quantum numbers $J^P=1^+$ the importance of mixing between different interpolator structures is highlighted. While almost physical sea-quark masses clearly improve agreement with experimental data, $D_s$ results in the $1^+$ and $0^+$ channels, where multiparticle thresholds are close to the state of interest, still show a substantial deviation from experiment. The role played by discretization effects, by finite volume effects and by an incomplete variational basis (providing only poor overlap with multiparticle states) will be a topic of future research.

\acknowledgments
We thank the PACS-CS collaboration for their gauge configurations and Martin L\"uscher for his DD-HMC software. The calculations were performed on computing clusters at TRIUMF and York University. We thank Sonia Bacca and Randy Lewis for making these resources available. This work is supported in part by the Natural Sciences and Engineering Research Council of Canada (NSERC).

\providecommand{\href}[2]{#2}\begingroup\raggedright\endgroup

\end{document}